**Modelling control strategies against Classical Swine Fever: influence of traders and markets using static and temporal networks in Ecuador.**

Alfredo Acosta [1], Nicolas Cespedes Cardenas[3], Cristian Imbacuan[2], Hartmut H.K. Lentz[4], Klaas Dietze[4], Marcos Amaku[1], Alexandra Burbano[2] Vitor S.P. Gonçalves[5], Fernando Ferreira[1]

1 Preventive Veterinary Medicine Department, School of Veterinary Medicine and Animal Science, University of São Paulo, SP, Brazil
2 Sanitary and phytosanitary control and regulation Agency-Agrocalidad, Ministry of Agriculture, Quito, Ecuador
3 Department of Population Health and Pathobiology, College of Veterinary Medicine, North Carolina State University, Raleigh, USA
4 Friedrich-Loeffler-Institut, Greifswald-Riems, Germany
5 Faculty of Agronomy and Veterinary Medicine, University of Brasilia, Brasilia, Brazil

Corresponding author: Alfredo Acosta, Preventive Veterinary Medicine Department, School of Veterinary Medicine and Animal Science. University of São Paulo. alfredojavier55@gmail.com, alfredoacosta@usp.br

**Abstract**

Pig farming in Ecuador represents an important economic and cultural sector. Recently, the National Veterinary Service has implemented individual identification of pigs, stricter movement control and mandatory vaccination against classical swine fever (CSF), which has been prevalent since 1940. The information registered since 2017 seems valuable for applying modelling techniques assessing the risk of dissemination of animal diseases and test control strategies. Social network analysis (SNA) was used to characterise the premises into risk categories, looking for better targeting surveillance activities and identifying higher risk targets. Network modelling was used to test the disease spread and control strategies over the network, using a Susceptible, Infected, Recovered, Susceptible SIRS model to compare a target selection of high-risk nodes against the current traditional surveillance. Finally, temporal network analysis revealed general trends in the network's evolution. The network contained 751,003 movements with 6 million pigs, excluding movements to slaughterhouses. A total of 165,593 premises were involved in network movements (144,118 farms, 138 industrials, 21,337 traders and 51 markets). On an annual average, 75% of the premises (124,976) received or sent up to one movement with an annual average of 1.5 pigs. In contrast, 0.01% of the premises (166) made up to 1,372 movements, with an annual average of 11,607 pigs. Simulations on the network resulted in the dissemination of CSF, reaching a mean prevalence of 29.93% without a control strategy. The random selection strategy reduced the prevalence to 24.4%, while target selection by risk reduced to 3.3%. When comparing the static representation of the network with the temporal counterpart, causal fidelity (c = 0.62) showed an overestimation of 38% in the number of transmission paths, even though it took an average of 4.39 steps to cross the network, these steps took approximately 233 days. The characterisation of the premises acting on the Ecuadorian network, based on risk emphasis, and the application of network modelling techniques can support the redesign of the surveillance system. Minor changes could enable the implementation of risk-based surveillance, targeting specifically the nodes at higher risk and considering special attention to markets and traders.

**Keywords**

Modelling; Network Analysis; Temporal networks; Classical swine fever; Target surveillance; Epidemiology.

## 1. Introduction

Pig farming in Ecuador represents an important economic sector, combined with traditional Andean culture of high consumption of pork meat (Benitez and Sanchez, 2001) (Valverde Lucio et al., 2021). Classical swine fever (CSF) was first recorded in Ecuador in 1940. In the last century, initiatives to control the disease were executed without success. Producers were responsible to vaccinate their pigs but lacked official coordination. In 2012, the National Veterinary Service (NVS) (known as Agrocalidad) established goals and coordinated activities to eradicate the disease through the National CSF Eradication project (Acosta and Vargas, 2012). Additionally, the mandatory continuous vaccination campaign against CSF started in 2017 with stricter movement control and automated tracing systems.

Classical swine fever is one of the most relevant viral diseases of swine, with a large public health impact is listed as a notifiable disease by the World organisation for animal health (OIE) (Ganges et al., 2020). The disease is highly contagious, affecting domestic and wild swine with multiple manifestations depending on various host and viral factors (OIE, 2021a); their first reports dated from 1822 in France and 1833 in the United States (Edwards et al., 2000). Currently, the disease is still present on all continents; however, OIE in 2021 recognised the free status of 38 countries. In Latin America, specifically in the Andean Community, the disease is still considered endemic, except for free zones in Colombia (large part of its central territory) and Ecuador (Galapagos islands) (OIE, 2021b). The continent has subscribed to a continental CSF eradication plan adding expectations to countries to progress in control efforts (Terán et al., 2004).

Social network analysis (SNA) is a popular technique capable of representing the patterns of interactions of complex systems (Newman, 2010) (James et al., 2009) (Wasserman et al., 1994) such as animal movements, involving large quantities of data that can be modelled as graphs of nodes connected by edges. Their study enables to comprehend the epidemiological patterns of infectious diseases in livestock, usually shaped by commercial and social structures (Dubé et al., 2009).

Another technique to study infectious diseases is mathematical modelling, which allows the use of individual-level knowledge of epidemiological factors to predict population-level epidemic dynamics (Keeling and Rohani, 2011); Currently, mathematical modelling is a pivotal tool for infectious disease studies showing how control strategies may change the dynamics and epidemiology of infectious diseases (Andraud and Rose, 2020) (Keeling and Danon, 2009) (Barongo et al., 2016). Consequently, network modelling techniques, a conjunction of social network analysis and traditional modelling, are convenient to capture contact patterns, simulate epidemics and test hypotheses (Craft, 2015).

Traditional network analysis has used static representations of movements over different time windows. Most times, edges are not continuously active over time; Therefore, the temporal structure changes the dynamics of the systems interacting through the network (Holme and Saramäki, 2012). Static networks ignore the temporal influences leading to an over or underestimation of the extent or even speed of the modelled outbreak (Masuda and Holme, 2013). Thus, temporal networks can improve our comprehension of epidemics than static networks.

Even though animal movements are key factors in between-farm disease transmission, modelling gives the opportunity to test different scenarios and strategies in silica. To date, network modelling techniques in animal movements have not been studied in Ecuador or in most of the Andean region. The objectives of this paper were to (1) characterise the premises according to risk criteria, (2) analyse the effect of different strategies over the network to control CSF, (3) identify key premises for practical surveillance and control interventions, and (4) analyse the temporal development of the network.

## 2. Material and methods

### 2.1 Origin of data

Registry of premises, vaccination against CSF and pig movements in Ecuador is mandatory. The vaccination campaign against CSF is the responsibility of officially certified swine producer associations. Pigs are registered, vaccinated and identified with official ear tags (vaccination against CSF is valid for 6 months). Then, producers (owners of any type of premise) issue their movement certificates through the online official system, selecting their vaccinated pigs based on their ear tag identification. Self-service emission of movement certificates started with industrial premises in 2014 (pilot tests) and later became mandatory to every premise in the country (including backyard). The NVS runs a supervision scheme using random selection of premises, checking the inventory of animals and clinical examination. Previous studies looking for how well premises that are connected in the network have been used to predict its risk of being infected and its potential as super-spreaders (Makau et al., 2021) (Alarcón et al., 2020).

### 2.1.1 Movement dataset

To analyse the pig movements, we accessed the official system called GUIA (Agrocalidad unified information manager) (https://guia.agrocalidad.gob.ec) and downloaded the movements from 2017 to 2019. The original database contained 1,212,008 movements with 10,419,708 pigs. We excluded 1.73% of the movements from the analysis, because of cancelled registries, cadastral issues, and not corresponding with the study period. The data used for the analysis were movement ID, date, origin and destiny premise, number of animals, type of premise, supervision ID and date of supervision. The type of premises as defined by the NVS are **Farm:** backyard premises; **Trader:** premises used to trade pigs, buying and selling animals, most frequently on the same day; **Industrial:** premises with commercial / multi-site intensive pig farming; and **Market:** authorised facilities where producers and traders market their pigs.

### 2.1.2 Surveillance data

We received data from outbreaks registered by the official surveillance system; the data used for the analysis was: number of reported suspicions and confirmed cases, number of positive results and population of pigs in farms, date of first clinical signs and date of resolution of the outbreak.

### 2.2 Descriptive analysis of movements

We analysed the pig movements as a static network or graph G = (V, E) which is a group of vertices V (nodes) and a group of edges E, every edge connecting a pair of nodes and constructs a directed network, where the edges account for the direction of the movement from one node to another (premise X sends pigs to premise Y). Movements to slaughterhouses were excluded because of epidemiological irrelevance (Guinat et al., 2016). Then, constructed networks of individual years and also of the entire time period. Networks were constructed using the igraph package in R. We created maps analysing the number of premises distributed in every parish and then calculated their annual average and the density of premises by $km^2$ to plot an annual representation according to their geographic distribution (Fig. 1).

Animal movement networks exhibit a high degree of heterogeneity, following the degree distribution P(k), which is the fraction of nodes in the network with degree k (scale-free networks), assuming that the degree distribution follows a power-law distribution where alpha is a scaling parameter, the exponent of the power law (Equation 1) (Ossada et al., 2013) (Newman, 2010).

$$P(k) \sim k^{(-\alpha)}$$

[1]

A high degree of heterogeneity is a characteristic of scale-free networks, presenting many nodes with a low number of contacts and a few nodes with a high number of contacts (Pareto Law). This factor influences the potential risk of spread of diseases (May and Lloyd, 2001). This property gives rise to the definition of 'super-spreaders', as infected hosts generating a disproportionately high number of secondary cases (Silk et al., 2017) (Lloyd-Smith et al., 2005) (Newman, 2008); We tested the Pareto rule and the cumulative distribution function to confirm these presumptions, calculating Pk = r/n (Pk is the fraction of nodes that have degree k or greater), sorting the degree of the nodes in descending order, numbering them from 1 to *n* (ranks of the nodes) divided by n (highest degree node), and plotting r/n as a function of degree k*i* (Newman, 2010).

### 2.2.1 Characterisation of premises according to risk criteria

We built a network of premises (slaughterhouses excluded) as nodes and edges as their shipments and calculated the degree and weighted degree (number of pigs). Then, the premises were classified according to their degree and grouped in deciles, showing their distribution by weighted degree.

For markets, we used the same procedure but classified their degree by quintiles to get 5 homogeneous groups. Finally, we analysed the characterisation and looked for categories with a higher risk to use in posterior simulations. We used box plots to obtain a better visualisation of the groups.

### 2.3 Modelling the dissemination and control of CSF spread in the static network

### 2.3.1 Model description

To model the disease spread on the network, we used the adjacency matrix (element *ij*). $A_{ij}$ = 1 if there is an edge between nodes *i* and *j,* otherwise $A_{ij}$ = 0. Then, defined the elements of the vector of infected nodes *I*. If node *i* is infected, then $I_i$ = 1, otherwise $I_i$ = 0. The result of the multiplication of the vector of infected nodes ( *I* ) by the adjacency matrix ( *A* ), is a vector ( *V* ), whose element *i* corresponds to the number of infected nodes that are connected to the node *i* and may transmit the infection (Ossada et al., 2013) (Equation 2).

$$V = I \cdot A$$
[2]

To simulate the disease spread on the network, we used a SIRS (susceptible-infected-recovered-susceptible) epidemic model according to the following scheme:

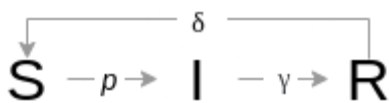

**S** represents a node susceptible to the disease, **I** represents an infected node, **R** represents a recovered node (effect of movement restriction), γ represents the number of time units in the simulation that the node will last infected and δ represents the number of time units that the node will last recovered from the disease. Then, calculated the probability (*p i*) of a susceptible node *i*, connected to *V i* infected vertices to become infected, using Equation 3:

$$p_i = 1 - \left(1 - \lambda\right)^{V_i}$$
[3]

where λ is the probability of disease spread, for each time step (days), and then updates the vectors of infected, susceptible and recovered nodes, considering the probability of disease spread. Infected nodes remained infected during a (γ) time randomly sampled between (*t*)**I**min and (*t*)**I**max. Recovered nodes remained immune during a time (δ) randomly sampled between (*t*)**R**min and (*t*)**R**max. Nodes under a strategy of movement restriction were not susceptible to infection (inmune) at random or target selection. The dissemination model through the network was originally described by (Ossada et al., 2013) and implemented in R.

## 2.3.2 Parameters used in the simulation

We used the 2019 network dataset for simulations. This network was more robust compared with the previous years of the study period, because in the first two years (2017 and 2018) the authority and stakeholders gradually adjusted to the system and procedures. To assess the spread of CSF through the network, we observed the change in the average daily prevalence, using the number of initially infected nodes, and the probability of infection. The infection probability used was 0.277, considering direct animal contact (high risk), for movement contact as simulated in the Netherlands (Mangen et al., 2002), considering the data from the 1997–1998 epidemic modelled by (Stegeman et al., 2002).

We used a prevalence of 5% of randomly infected farms to start the simulation (we analysed the official records of outbreaks, looking for herd prevalence, and also consulted local experts because of lack of published information about national prevalence). The local experts were 3 specialist of the national CSF official control project in the headquarters in Quito. To define the minimum and maximum infection times of the premises, we analysed the historic outbreak length, from the onset of clinical signs to the conclusion, according to official records, considering the first and third quartiles of the number of days.

In the same way, to determine the minimum and maximum recovered times, we considered that an infected farm would be mandatorily empty at least 30 days after the closure of the event (~2 CSF incubation periods of 15 days) until 6 months (local expert opinion). To illustrate this, backyard producers, after confronting an outbreak, usually change their activity to agriculture or to rising another small species such as chickens or guinea pigs (Encyclopedia Britannica, 2021) (National Research Council, 1991), species with short production cycles, and then start again with pigs after approximately 6 months (Table 1).

Table 1. Initial conditions, parameters and values of the network model in Ecuador.

| | Value | Description | Reference |
|---|---|---|---|
| **Initial conditions:** | | | |
| Premises (nodes) | 93,707 | Number of nodes present in 2019 network | Dataset |
| Initial infected nodes | 4,685 | 5% Initial | Agrocalidad 2018. Surveillance system. |
| Movements (edges) | 314,38 5 | Number of movements present in 2019 network | Dataset |
| **Parameters:** | | | |
| Prob (p) | 0.277 | Probability of disease spread from an infected to a susceptible node | (Mangen et al., 2002) |
| $tI$min (d) | 42 | Minimum time a node remains infected (day) | Agrocalidad 2018. Surveillance system. |
| $tI$max (d) | 71 | Maximum time a node remains infected (day) | Agrocalidad 2018. Surveillance system. |
| $tR$min (d) | 30 | Minimum time a node remains recovered (day) | Agrocalidad 2018. Surveillance system. |
| $tR$max (d) | 180 | Maximum time a node remains recovered (day) | Agrocalidad 2018. Surveillance system. |
| Tsim (d) | 365 | Simulation time | |
| **Random selection conditions:** | | | |
| Random selection strategies | 2.25, 3.14, | Percentages of premises under supervision by the NVS | Dataset |
| Random selection strategies | 5, 10, 25, 50 | Percentages of premises under random selection for comparison | Dataset |
| **Target selection conditions:** | | | |
| Centrality metrics (Degree) | 10 | Nodes with highest degree individual metric | (Cardenas et al., 2019) |
| Centrality metrics (pagerank) | 10 | Nodes with highest pagerank individual metric | Dataset |
| Type of nodes | 10 | Markets | (Motta et al., 2017) |
| Type of nodes | 10 | Traders | |

| Geographic distribution | 2,300 | 100 highest degree premises stratified by geographic locations (23 provinces) | Dataset |
| Risk characterisation | 10 | Upper quintile of markets. | Dataset |
| Risk characterisation | 2 | Upper quintile markets (AsoganSD*, Saquisili*. | Dataset |

(p)= probability, (d)= days, *Names of markets in spanish.

### 2.3.3 Random selection strategy

We analysed the movement inspection activities against CSF. In 2019, the NVS inspected 2.25% of the movements in origin or destination and 3.14% in road operatives. The random selection strategy used a percentage of nodes to implement a disease control of the dissemination of CSF. We set those percentages at 2% and 3%, emulating the random control in 2019, and setting additional values to 5%, 10%, 25% and 50% for comparison. This percentage considered the ($t$) min and ($t$) max when the nodes (premises) remained immune or recovered (Table 1).

### 2.3.4 Target selection strategy

We selected two centrality metrics to test the target selection strategies, degree and pagerank. The last is the popular algorithm used by google, derived from Katz centrality where the centrality of my network neighbours is proportional to their centrality divided by their out-degree (Newman, 2010). Furthermore, a large pagerank shows the importance of connecting different components of the network (Farine and Whitehead, 2015). Degree is the sum of ingoing and outgoing movements, in-degree is the sum of individual sources providing animals to a specific premise, and out-degree is the number of individual recipients getting animals from specific premises (Wasserman et al., 1994).

The target selection strategy considered the selection of premises with higher individual metrics and tested specific strategies that the NVS could plausibly apply with the available resources, infrastructure and data. Finally, we considered (1) the highest degree and pagerank individual metrics, (2) the type of nodes: traders and markets, (3) geographic location: distributed in each premise and similar to the NVS random strategy and (4) risk characterisation (above).

To evaluate the effect of the strategies, we analysed the prevalence of infection over the network, looking at the individual states of the nodes as infected or not infected on every simulation day, evaluating the annual prevalence and plotting the curves of infection (Table 1).

### 2.4 Temporal network analysis

A temporal network is a graph in which each edge is time stamped specifying the time at which its endpoints communicate (Kempe et al., 2002). A temporal network (graph) $G = (V, E)$, where $V$ is a set of nodes and $E$ is a set of temporal edges. In temporal networks, it is possible to build an accessibility graph of a network by consecutively adding paths of growing length (unfolding) to get information about characteristic time scales and the distribution of the shortest path durations (Lentz et al., 2013). The unfolding accessibility method also extracts temporal information about a potential epidemic outbreak, emulating a Susceptible-Infected (SI) spreading process on the temporal network (Lentz et al., 2016).

We compute the causal fidelity ($c$), defined as the fraction of paths in the static network that are also present in the temporal network. In addition, the causal error ($e$) is defined as the reciprocal causal fidelity (Equation 4) and is used to quantify the factor by which a static network overestimates network connectivity.

$$c = \frac{\sum_{ij} p_{ij}}{\sum_{ij} P_{ij}} = \frac{\rho(p)}{\rho(P)} = \frac{1}{e} \qquad [4]$$

where $\rho(p)$ is the total number of casual paths in the temporal network and $\rho(P)$ is the total number of paths in the static counterpart, an edge $(i, j)$. The 'shortest path duration' from node *i* to *j* is the connection where the time in the causal path is minimal. The derivative of the number of paths regarding time gives the shortest path duration distribution, which we used to detect characteristic time scales of the network.

## 2.5 Software

We performed statistical analysis using R version 3.6.3 (R Core Team, 2020) and R packages: gdata (Warnes et al., 2017), ggmap (Kahle and Wickham, 2013), ggsn (Baquero, 2019), igraph (Csardi and Nepusz, 2006), Matrix (Bates and Maechler, 2021), raster (Hijmans, 2020), rgdal (Bivand et al., 2020), sp (Bivand et al., 2013) and tidyverse (Wickham et al., 2019). Temporal analysis used Python3 libraries: scipy (Virtanen et al., 2020), numpy (Harris et al., 2020) and networkx (Hagberg et al., 2008).

## 3. RESULTS

### 3.1 Descriptive analysis

Ecuador in northwestern South America, crossed by the equatorial line, has 4 natural regions: the Highlands, Littoral, Amazon and Insular regions (Galapagos), and its land area is 248,360 km$^2$. The Administrative division in Ecuador has three levels: Province (24), Canton (221) and Parish (1,032). Eight point five percent of parishes did not register any movements in the study period, attributed to sub notification, registry errors and lack of swine population, especially in the Amazon region. The average number of premises by parish was 58.51 (min=1, Q1=3, Q3=38, max=2,689), while, the average density (premises/km$^2$) was 1.59 (min=1, Q1=1, Q3=1, max=30), distributed in 944 (91.5%) of the parishes. Premises were more concentrated in the North and Central Andean highlands and also in the western lowlands (Fig. 1).

**Fig. 1.** Study area showing the spatial distribution and histogram of premises registered in each parish from 2017 to 2019 in Ecuador. a) Annual average number of premises. b) Annual average

density (premises/km$^2$) of premises. Polygon lines represent parish limits (grey lines). White polygons represent parishes without registered premises.

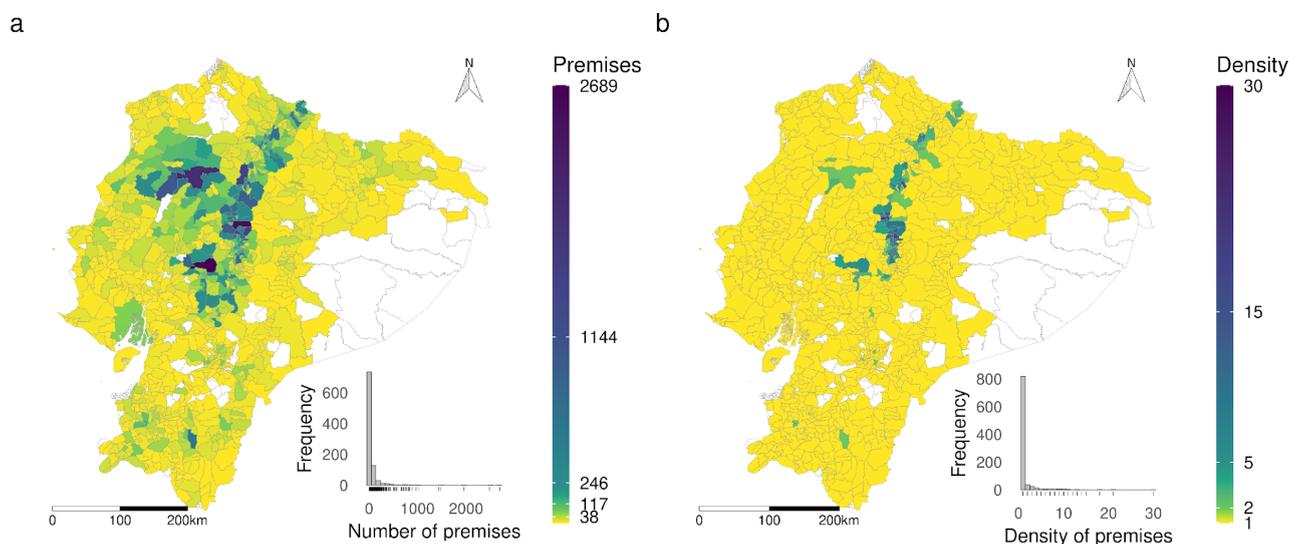

The complete dataset contained 1,190,991 movements with 9 million pigs. Excluding movements to slaughterhouses, the final working network involved 751.003 movements with 6 million pigs, and the total number of animals moved was 1,783,745, 2,147,794, and 2,475,853 in 2017, 2018 and 2019, respectively. The origin of movements was 57% by farms, 34.3% by markets, 4.4% by traders and 4.3% by industrials. Looking at the number of animals in the network, industrials moved 47.4% of the pigs, followed by 32.8% from farms, 16.4% by markets and 3.48% by traders. The annual average path length was 4.39 (Fig. 2).

**Fig. 2.** Movements according to the type of premises in Ecuador. 2017–2019. a) Number of movements. b) Number of animals.

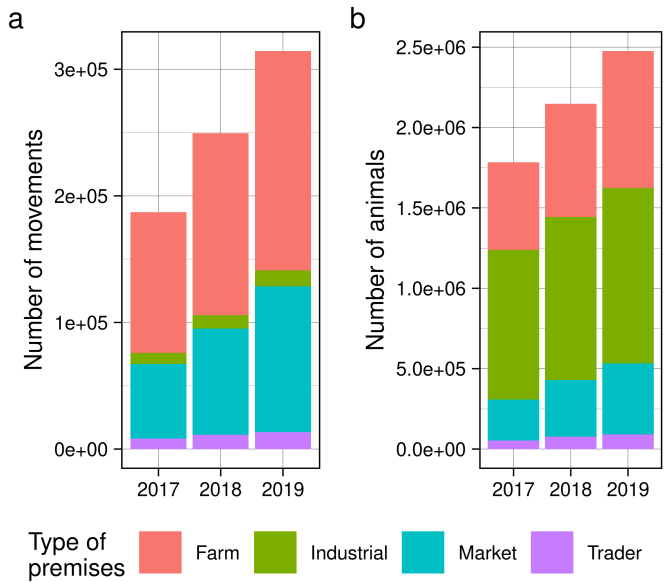

The Ecuadorian network fulfils the scale-free network assumption, as large numbers of nodes have a low number of contacts, and few have a high number. Their cumulative distribution function with straight-line behaviour for the degree of nodes suggests that follows a heavy tail bimodal distribution (bump on right) (Fig. 3).

**Fig. 3.** Cumulative distribution function. Heavy tail distribution in the Ecuadorian network.

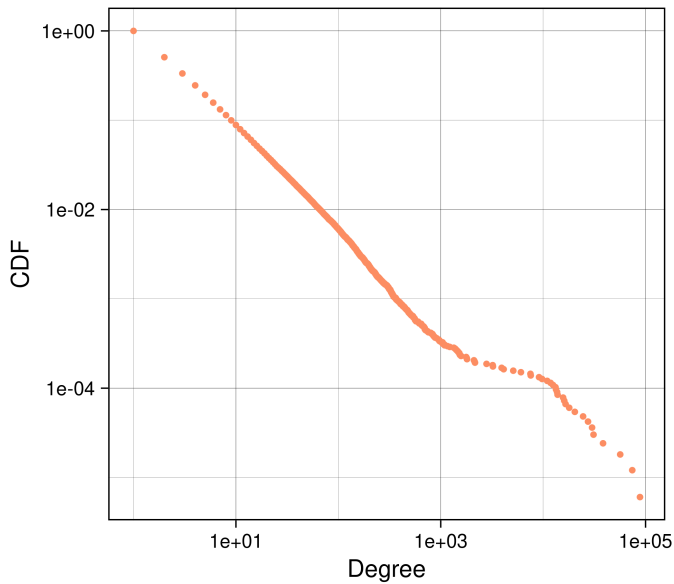

## 3.2 Characterisation of premises

The veterinary service registered 165,593 premises from 2017 to 2019; on an annual average, 75% (124,976), received and sent up to one movement with an average of 1.4 pigs. In contrast, 0.1% of the premises (166) made up to 1,372 movements, with a mean of 11,607 pigs. The median of weighted degree up to 99.5th percentile was 5.16 pigs. (Table 2).

**Table 2.** Network centrality values and characteristics of premises grouped by percentiles of their degree and weighted degree (pigs) 2017-2019 in Ecuador. Yearly average values of the study period.

| Percentile | Number of premises | Percentage | Degree | | Weighted degree (pigs) | | |
| --- | --- | --- | --- | --- | --- | --- | --- |
| | | | Mean (Min-Max) | Median | Mean | Max | Median (Q1–Q3) |
| 75th | 124,976 | 75.5% | 0.49 (0.33–1) | 0.33 | 1.4 | 500.67 | 0.17 (0.67–0.84) |
| 90th | 24,159 | 14.6% | 1.78 (1.33–2.67) | 1.67 | 6.8 | 500.00 | 1.34 (4.33–4.17) |
| 99.5th | 15,638 | 9.4% | 7.56 (3 –36.7) | 5.00 | 45.2 | 9,347.33 | 5.16 (20.33–22.5) |
| 99.9th | 654 | 0.4% | 57.38 (38–105.6) | 52.33 | 594.9 | 18,938.33 | 96.46 (343,5–309.5) |
| 100th | 166 | 0.1% | 201.2 (107–1,372) | 152.33 | 11,607.5 | 245,370.33 | 327.34 (1,212.33–1,824.34) |

The premises of the 75th percentile were 86% farms (107,483), 14% traders (17,483) and 0.001% industrials (13), representing most of the premises. In contrast, in the upper percentile were 48% farms (80), 25% traders (41), 27% industrials (45); also, the median degree of the 45 industrial premises was 191,33 and their median weighted degree was 29,688.3. Traders of the upper percentile had an annual median degree of 142 and weighted degree of 895 pigs. This means that they made almost 3 movements with 17 pigs per week, it is common to see them buying and selling pigs on farms or between markets all around the country. Because of their activity, traders in the two upper percentiles could represent a high risk for disease spread (super-spreaders), specifically because they connect different premises and locations around the country (Table 3). (Fig. 4).

**Table 3.** Categorisation of type of premises by percentiles of their mean degree and weighted degree (pigs) 2017-2019 in Ecuador. Yearly average values of the study period.

| Percentile | Type of farm | Premises | Degree | | Weighted degree (pigs) | | |
|---|---|---|---|---|---|---|---|
| | | | Mean | Median | Mean | Max | Median (Q1–Q3) |
| 75th | Farm | 107,483 | 0.50 | 0.33 | 1.50 | 500.67 | 0.67 (0.165–0.84) |
| 75th | Trader | 17,480 | 0.44 | 0.33 | 1.22 | 80.00 | 0.67 (0.165–0.5) |
| 75th | Industrial | 13 | 0.54 | 0.33 | 24.23 | 200.00 | 3.67 (0.335–10.67) |
| 90th | Farm | 22,257 | 1.79 | 1.67 | 6.78 | 500.00 | 4.33 (1.335–4) |
| 90th | Trader | 1,894 | 1.76 | 1.67 | 7.73 | 100.00 | 4.33 (1.335–4.335) |
| 90th | Industrial | 8 | 2.08 | 2.00 | 93.46 | 419.00 | 36.33 (14.165–36.71) |
| 99.5th | Farm | 13,873 | 7.32 | 5.00 | 40.88 | 9,347.33 | 20 (5–21.5) |
| 99.5th | Trader | 1,726 | 9.40 | 6.00 | 65.18 | 1,532.33 | 24.83 (5.665–32.165) |
| 99.5th | Industrial | 39 | 13.66 | 9.00 | 718.14 | 8,715.67 | 255.67 (40.585–239.665) |
| 99.9th | Farm | 425 | 56.21 | 51.33 | 407.85 | 2,685.33 | 291.67 (86.5–257.335) |
| 99.9th | Trader | 196 | 58.34 | 52.67 | 560.93 | 6,186.67 | 388.5 (109.335–361.54) |
| 99.9th | Industrial | 33 | 66.75 | 66.33 | 3,206.13 | 18,938.33 | 1800.33 (503.335–1913.165) |
| 100th | Farm | 80 | 196.47 | 146.83 | 1,395.53 | 8,989.00 | 804.5 (229.085–690.585) |
| 100th | Trader | 41 | 152.17 | 142.00 | 1,415.33 | 4,222.00 | 895 (337.5–938) |
| 100th | Industrial | 45 | 254.58 | 191.33 | 39,048.41 | 245,370.33 | 29,688.33 (2,706.34–22,256.17) |

The type of premise did not correspond with the number of movements and official definition most times, because of registry errors or lack of clear definition of the procedures and system classification. For example, premises defined as mainly backyard, are presented in the upper percentiles, moving hundreds/thousands of animals (Fig. 4).

**Fig. 4.** Premise box plots of weighted degree (number of pigs) (log scale) classified by percentiles of their degree. Distributions grouped by type of premise in the Ecuador annual average of the study period. (Labels with the number of premises in each group and percentiles).

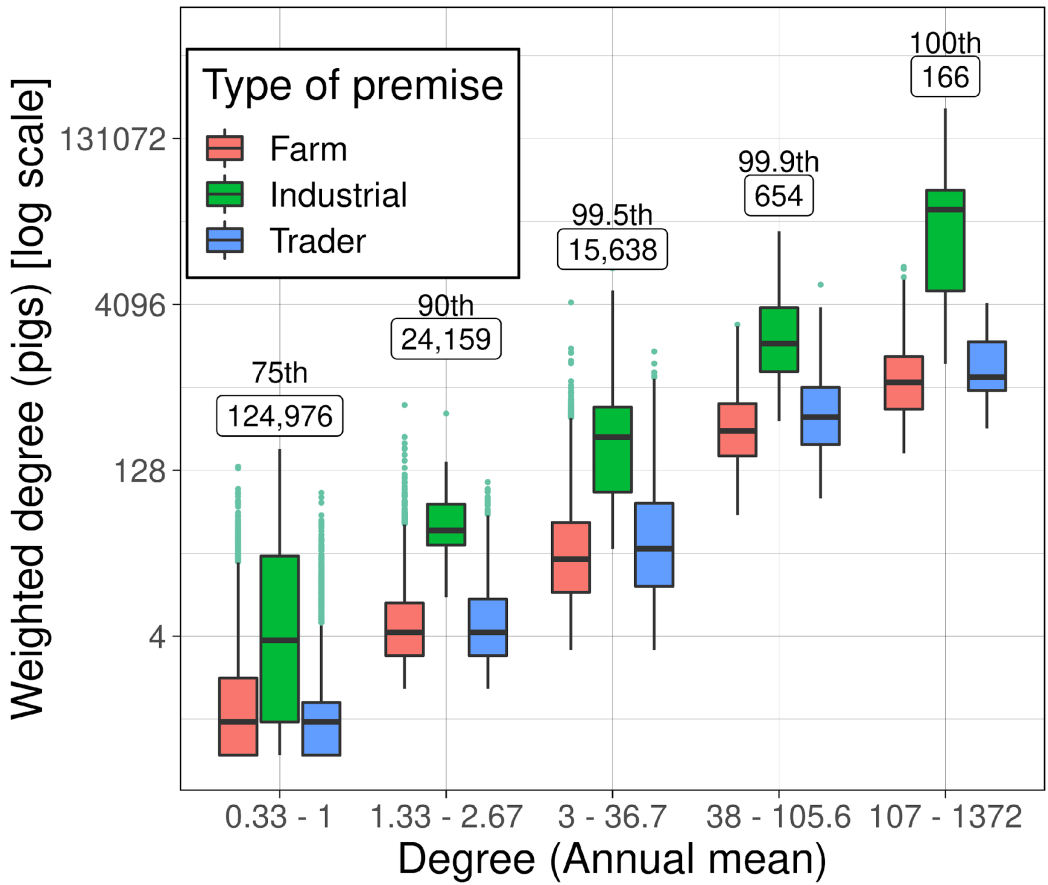

Markets are principal hubs on the Ecuadorian network. It was possible to identify and categorise markets with low degrees such as 'Canar' in the first quintile and enormous markets such as 'AsoganSD' in the fifth quintile. The average degree by quintile (QU) was QU1 = 83.6, QU2 = 456, QU3 = 1,948, QU4 = 4,610 and QU5 = 13,628. Comparing the average degree between the first and fifth quintiles, the last was 162 times greater in the number of movements and 151 times greater in the average number of animals (Q1 = 395, Q5 = 59,684) (Fig. 5).

Fig. 5. Market box plots of weighted animal degree distribution, classified by quintile of degree in Ecuador (Markets' original names in Spanish) Average annual values in the study period.

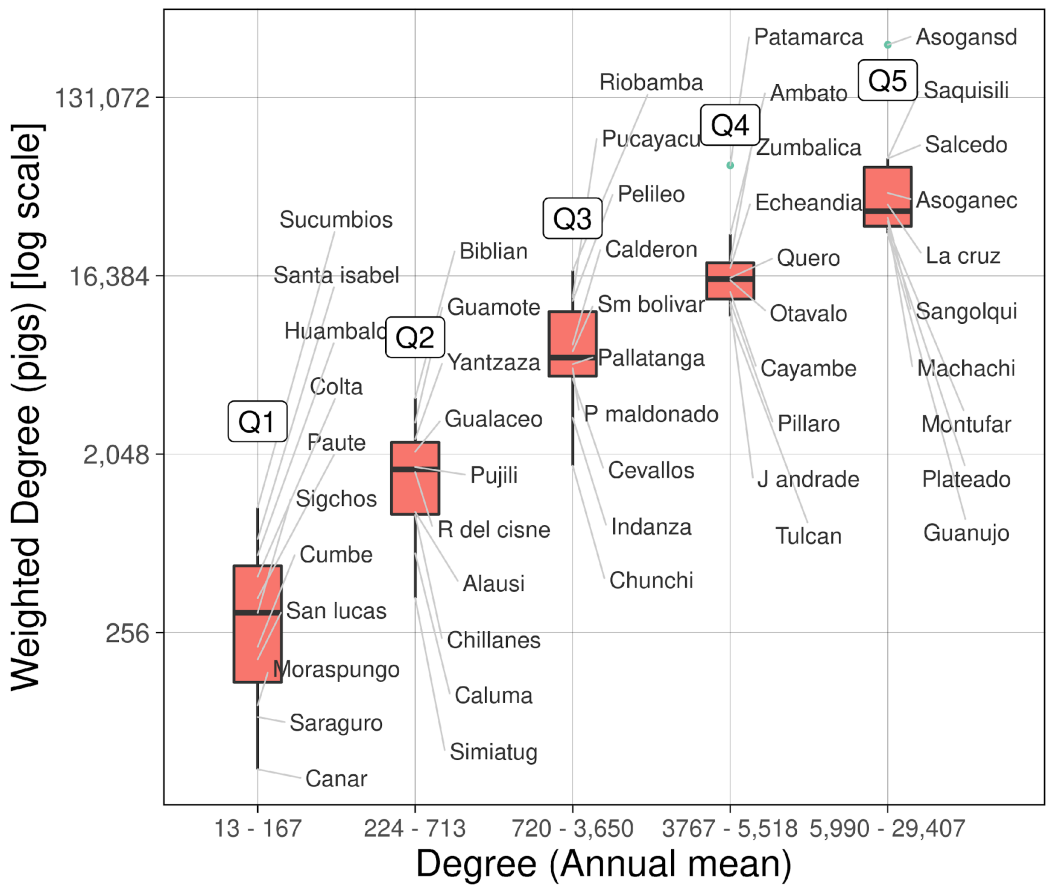

Markets in the upper quintile had an annual average weighted degree of 59,684 pigs; these markets received between 500 and 4,600 animals each week (some of them two days per week), generating thousands of high-risk contacts in a few hours of activities (04H00–15H00). This suggests that the larger the market is, the higher the potential risk for spreading diseases. Posterior simulations in this paper will use some of these markets to evaluate their role in disease spread over the network. Centrality metrics (degree, pagerank) of the 51 markets are available in supplementary materials SM1 and SM2. Markets are geographically distributed mostly in the highlands (Andean zone) (Fig. 6), with only 3 on the littoral zone and 2 on the Amazon.

Fig. 6. Representation of markets on the physical map of Ecuador. Mountain regions characterised the highlands going from north to south, dividing the littoral region to the west up to the Pacific Ocean and Amazon region to the east. Markets are represented as circles on the map, coloured by their quintile group. Size of groups according to their importance (risk) on the network (not scaled).
Names of the markets of the higher quintile on the map.

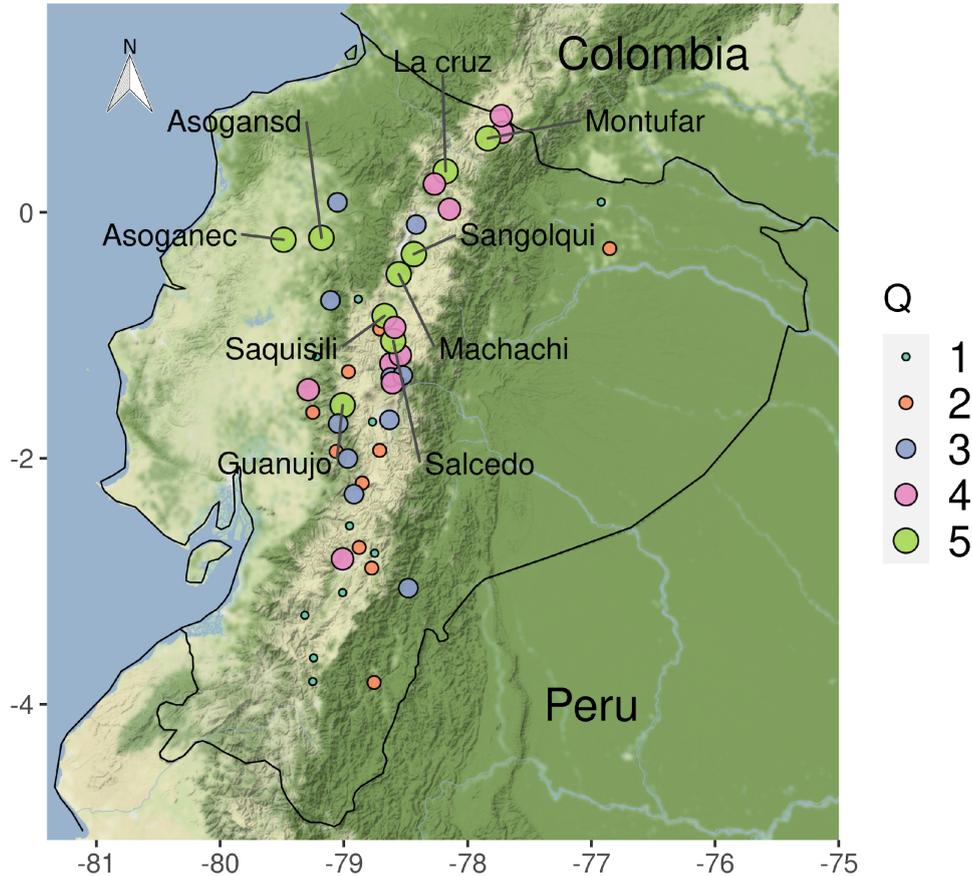

The premises that maintained their activity over the three-year period were 13,727 farms (9.52%),
94 industrials (67.63%), 46 markets (86,79%) and 1.546 traders (5.66%), evidencing a low fidelity
on the network. Most of the premises were active on the network for one year, but not the next.
Self-consumption of pigs and irregular movements are common practices in rural areas that could
explain this behaviour.

### 3.3 Dissemination of CSF spread through the Network.

3.3.1 Outbreak analysis

In 2019, the surveillance system confirmed 35 CSF outbreaks from 167 reported suspicious.
Considering the cumulative number of animals from the population at passive surveillance (11,176

pigs) and the animals affected (death and culled) by the disease (435 pigs), resulted in a prevalence of 3.89%. The local experts considered this prevalence and the limitations on sensitivity and representativeness of the surveillance system, defining a 5% starting prevalence on the simulations. The average number of days of the outbreak length from identification of the first sick animal until the closure of the outbreak was 57.52 days (min = 10, Q1 = 42, median = 52.50, mean = 57.52, Q3 = 71.25, max = 193). 120 of the suspicions premises were backyard (71.85%) and 43 commercial (25.15%). Only 1 of the 35 outbreaks were fully vaccinated against CSF. 72% of the outbreaks did not register any movements. 85.72% of the outbreaks were on the 75[th] degree percentile (1–3) and 14.29% on the 99.5[th] percentile (3–36). Irregular movements, biosecurity failures and lack of immunisation were registered as the major causes of infection according to the reports. The information from databases, vaccination, movement and surveillance were not fully integrated, representing one limitation of the study.

3.3.2 Modelling results

When modelling the spread of CSF through the network (without a control strategy), the mean infection prevalence was 29.93%. The epidemic curve presented two peaks, the first on Day 43, with 52.8% infected nodes, and the second on Day 305, with 30.56% infected nodes (Fig. 7a).

A random selection strategy reduced the prevalence of CSF by 18.26% when controlling 2,108 nodes (2.2%), compared with the prevalence resulting from free dissemination. Higher percentages of nodes under movement restriction also reduced infection over the network (Fig. 7a). Meanwhile, the number of nodes was very high; we observed reductions of 53% controlling 23,000 nodes, and 80% controlling 46,000 nodes. However, implementing these strategies is very unlikely (Table 3) (Fig. 7b).

**Fig. 7.** Simulated daily prevalence of CSF in the swine network over a calendar year. a) Prevalence over time of random control strategies (percentage of random nodes under control). b) Box plots of the simulated prevalence of every strategy (mean prevalence value on box plots).

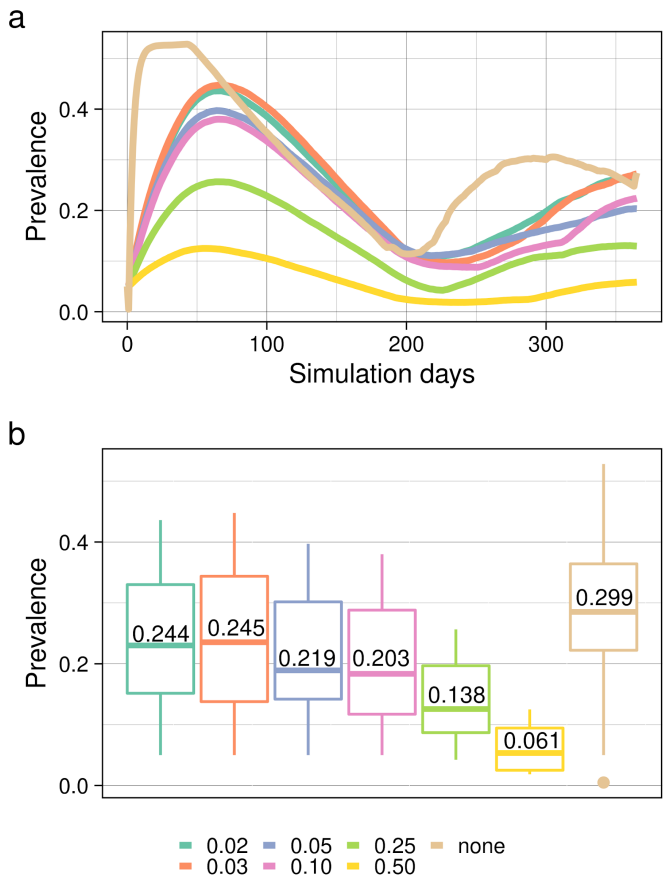

When quantifying the effect on prevalence reduction using target selection, controlling 10 premises (higher all-degree values) showed a prevalence reduction of 87.91%. Seven of ten nodes with higher pagerank were coincident with higher degree (then we selected the next ones); the results showed a prevalence reduction a little superior than degree (88.31%).

The target selection of 53 markets resulted in a prevalence reduction of 88.52%. Likewise, target selection of the top 10 national traders resulted in a prevalence reduction of 86.21%, displaying their relevance to disease transmission over network dynamics (Table 4). The simulation for traders behaved similarly to markets simulations based on pagerank and degree selection. The highest prevalence (0.090) was on simulation Day 56 for markets and Day 63 for traders (Fig. 8a).

**Table 4.** Simulation results of random and target selection strategies against CSF in Ecuador.

| | Details of strategy | Nodes under movement restriction [a] | Mean prevalence [b] | Prevalence reduction (%) | SD | MD | Range (min-max) |
|---|---|---|---|---|---|---|---|
| | None | 0 | 0.2993 | - | 0.12 | 0.29 | (0–0.53) |
| Random selection (% of nodes removal) | 2.25% | 2,108 | 0.2443 | 18.36% | 0.11 | 0.23 | (0.05–0.44) |
| | 3.14% | 2,904 | 0.2455 | 17.97% | 0.12 | 0.24 | (0.05–0.45) |
| | 5.00% | 4,985 | 0.2194 | 26.69% | 0.10 | 0.19 | (0.05–0.4) |
| | 10.00% | 9,370 | 0.2033 | 32.09% | 0.10 | 0.18 | (0.05–0.38) |
| | 25.00% | 23,426 | 0.1388 | 53.64% | 0.07 | 0.13 | (0.04–0.26) |
| | 50.00% | 46,853 | 0.0611 | 79.57% | 0.04 | 0.05 | (0.02–0.12) |
| Target selection (Strategies of nodes removal) | 2300 nodes (top 100) [c] | 2,300 | 0.0339 | 88.67% | 0.02 | 0.02 | (0.01–0.07) |
| | 53 markets | 53 | 0.0343 | 88.53% | 0.02 | 0.02 | (0.01–0.07) |
| | Pagerank (top 10) | 10 | 0.0350 | 88.31% | 0.02 | 0.02 | (0.01–0.07) |
| | Degree (top 10) | 10 | 0.0362 | 87.92% | 0.03 | 0.03 | (0.01–0.08) |
| | Traders (top 10) | 10 | 0.0413 | 86.21% | 0.03 | 0.03 | (0.01–0.09) |
| | AsoganSD [d] | 1 | 0.0416 | 86.11% | 0.03 | 0.03 | (0.01–0.09) |
| | Saquisili [d] | 1 | 0.0417 | 86.06% | 0.03 | 0.03 | (0.01–0.09) |

[a] Nodes under a strategy of movement restriction were not susceptible to infection. [b] Reduction compared with the simulated prevalence without control strategy. [c] Top 100 in each province, only farms considered. [d] Name of the markets.

When selecting the top 100 nodes in every province (2,300 nodes), the highest prevalence reduction was 88.67%. This strategy could be worth mentioning because of the easy applicability by official veterinarians (100 nodes in each province or 2 nodes in each parish, approximately).

Target selection of the two major swine markets (Saquisili and AsoganSD) resulted in a reduction of 86% of the mean prevalence; This strategy was historically executed in Ecuador to control CSF epidemics, however it was very hard to implement because of the affectation of thousands of producers and traders. Markets AsoganSD and Saquisili (Fig. 8b) showed almost equal dominant influence on disease spread, which contradicts the popular perception of AsoganSD as the most important in the country.

**Fig. 8.** Simulated daily prevalence of CSF in the swine network. a) Epidemic curves for target selection strategy simulations over 1 year. b) Box plots of daily prevalence in every simulation.

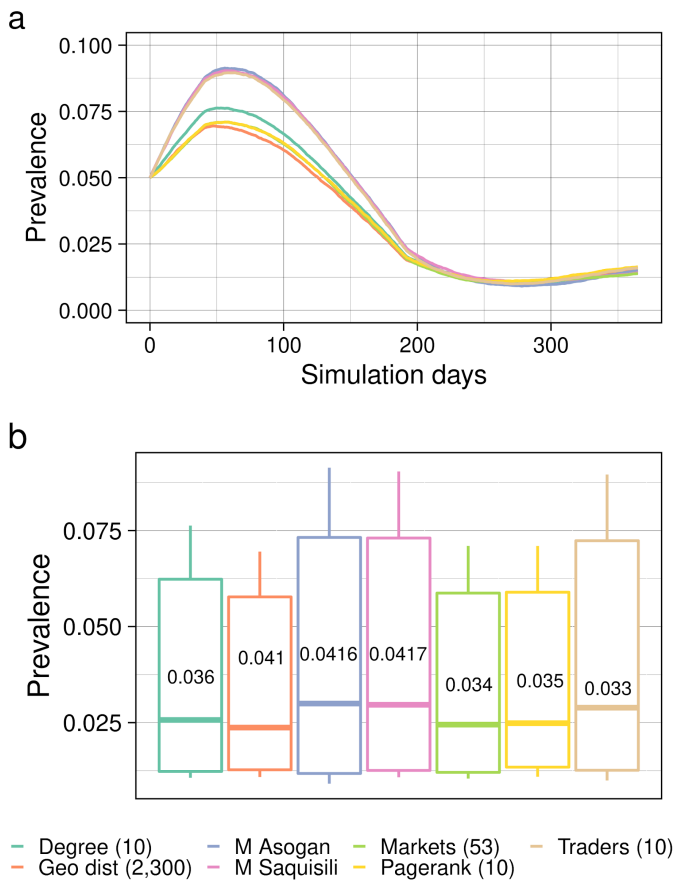

### 3.4 Temporal analysis.

The complete network had 165,648 nodes at time resolution of 1095 days, and pig trade movements were present every day of the study period. (Table 3). We observed 5,976,424,830 causal paths and 9,648,627,330 static paths on the network (5,457,922 causal paths every day); thus the causal fidelity was c = 0.62, representing that 38% of the paths considered on the static network do not exist. The causal error ($e$ =1.61) showed that considering the worst-case scenario over the entire study period, the static network would overestimate the size of an outbreak of disease by a factor or 1.61 (Table 5).

**Table 5.** Temporal network description.

| Data set | Size (nodes) | Edges | Causal paths (temporal) | Static paths | Causal fidelity | Error | Time resolution (days) |
|---|---|---|---|---|---|---|---|
| 2017–2019 | 165,647 | 750,003 | 5,976,424,830 | 9,648,627,330 | 0.6194 | 1.614 | 1,095 |
| 2017 | 59,689 | 172,69 | 572,445,956 | 1,036,256,373 | 0.5524 | 1.810 | 364 |
| 2018 | 75,221 | 232,708 | 974,845,227 | 1,778,922,594 | 0.5479 | 1.824 | 365 |
| 2019 | 93,707 | 292,257 | 1,592,439,715 | 2,913,437,409 | 0.5465 | 1.829 | 365 |

We represented the cumulative path (black lines) and shortest path duration (red lines) of the network in Fig. 9, and the shortest path duration shows saturation peaks approximately at 200, 600 and 900 days.

**Fig. 9.** Shortest and cumulative paths over the network in the Ecuadorian pig production system from in the study period.

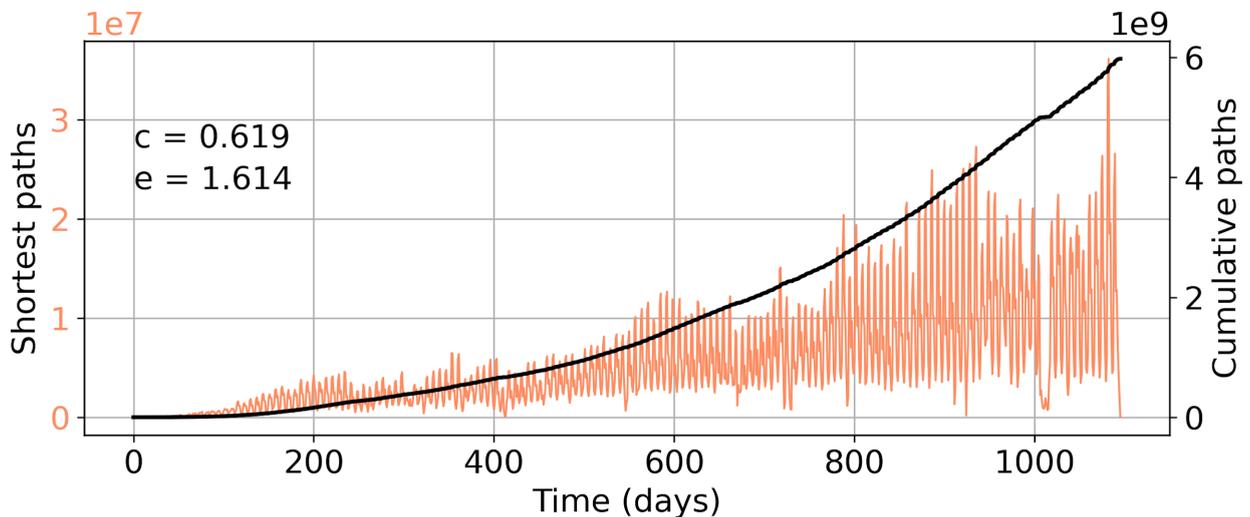

The individual annual analysis of the networks shows the partial saturation of the path density at 234 days for 2017 (Fig. 10a), 262 days in 2018 (Fig. 10b) and 205 days in 2019 (Fig. 10c) and a short saturation around the last two weeks of every year (353 days). These peaks, 233 days on average, could represent a typical time scale for infection spread, determined by the lifetime of livestock pigs, at approximately 6–8 months. This time scale is inherent to the network information (as a substrate for spreading) and does not depend on any disease parameter, as usually done with modelling techniques. Additionally, the most likely time for a disease to reach a random node in the network is around 234 days. On average 4.39 steps (average path length) are needed to cross the network, but these steps take in most cases 233 days (Fig.10).

**Fig. 10.** Shortest and cumulative paths over yearly networks in the Ecuadorian pig production system from 2017 to 2019.

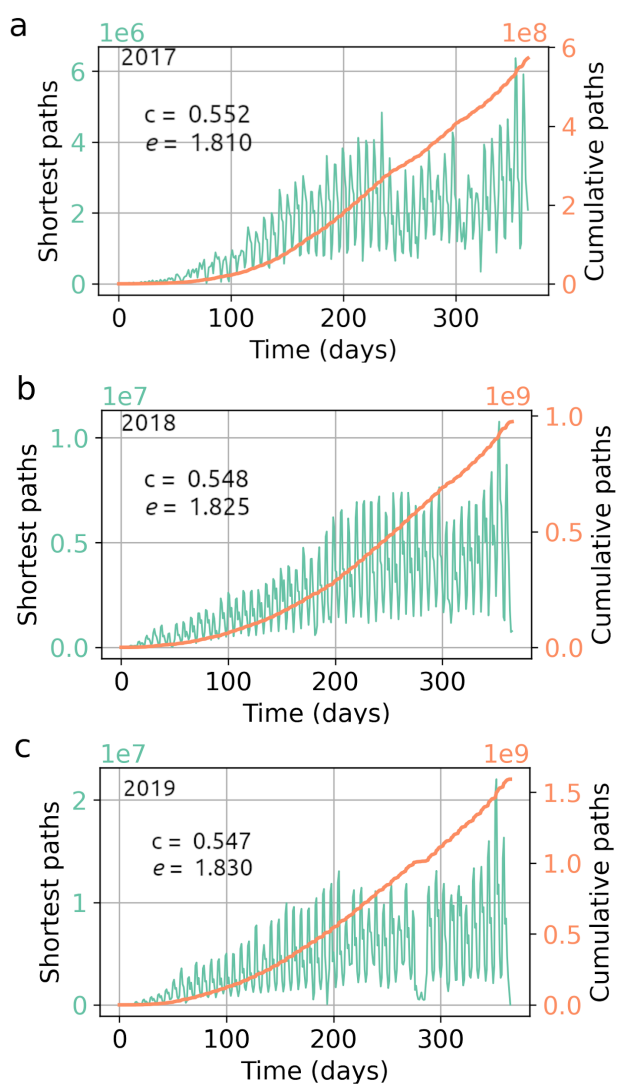

## 4. Discussion

Pig premises in Ecuador are mainly backyards (64.91%), taking part in the network with up to 1 ingoing or outgoing movement every year, and a few industrial farms (0.03%) taking part with thousands of movements; indeed, the global pig sector shares this dichotomy, between the production type and biosecurity level (Beltran-Alcrudo et al., 2019) (Baron et al., 2020). Meanwhile, in between these dichotomous extremes, premises with high capacity of movement and low level of biosecurity are at higher risks for the dissemination of infectious diseases. Outbreak data showed that most of them (85.72%) were in the 75th percentile of movement category. The supervision of premises according to their risk category could be an important step in implementing better surveillance and control strategies against animal diseases in Ecuador.

The type of premises categories sometimes do not correspond with their number of movements, animals and definitions, which happens especially with traders and because there is no systematic supervision or analysis of the declared category. However, automatic classification techniques (Koeppel et al., 2018) (Frössling et al., 2014) based on network parameters could be an ideal solution to improve the overall structure of the database and to help define target selection strategies of premises focused on risk categories. The NVS could focus supervision on specific combinations of the characteristics of those groups, check and change the requirements to reclassify the premises. It could also improve the information system to manage these changes by official veterinarians.

Animal markets are well known for being highly connected components of the network (Bigras-Poulin et al., 2006) (Robinson and Christley, 2007) (Baron et al., 2020). In Ecuador, pig markets are geographically concentrated in the central highlands with large backyard populations, related to a higher risk of presentation of CSF (Acosta et al., 2019). Markets behave like important hubs where the application of specific procedures and sanitary policies are needed. Therefore, to make an efficient control of CSF or any other disease, controlling the higher risk category (Q5 with 10 markets) resulted as efficient in prevalence reduction as controlling 51 of them. Markets could be risk categorised according to centrality metrics, as presented here, focusing the sanitary efforts and limited resources.

Considering the reductions in the simulated average prevalence, controlling 10 premises with the highest degree or pagerank values was more efficient in reducing prevalence than random control of 50% of the premises, illustrating how important the specific selection of high-risk premises is, and how disproportionately important these premises are from an epidemiological perspective (Knific et al., 2020). Our findings are consistent with those by (Cardenas et al., 2021) in Brazil, where the removal of premises based on their degree compared to random removal reduced the potential for transmission of infectious pathogens.

Every target selection strategy based on network centrality metrics evidenced better prevalence reduction than random selection. The result of the reassignment and reduction of the random premises selected by the NVS, represented a significant gain of 75% in prevalence reduction, controlling 20% fewer premises, evidencing the power of a more efficient control strategy.

This is the first attempt to measure the influence of traders. The selection of the top 10 traders reduced 88% of the simulated prevalence. This represents a surprising and important result because of the actual few regulatory actions on them, compared with markets that have a specific regulatory frame associated with their permanent infrastructure. However, considering their disease spreading potential, both types of premises deserve equal attention. In Colombia, pig traders also share the intense activities of buying and selling animals between municipalities without sanitary controls (Pineda et al., 2020). This reassures the local and regional importance of traders and motivates the necessity to comprehend better their influence on disease spread.

The majority (97%) of outbreaks in 2019 occurred in non-vaccinated animals, regardless significant vaccination efforts of the NVS in the last few years. Acknowledging the challenges in obtaining higher vaccination coverage, by analysing the influence on disease spread, the NVS should emphasize the vaccination coverage of the identified disease drivers of the supply chain (higher risk hubs). Nowadays only vaccinated pigs (individually identified by ear tags) are allowed to enter markets, preventing spread events mitigating the risk of disease spread. Additionally, a mechanism to maintain the individual traceability on markets has been established. However, strategies aiming at traders are still missing, despite the fact that many of them could become disease-spread hubs. Adapted disease control strategies should incorporate this knowledge gain on the traders role in disease spread within the pig commercialization structure.

Temporal analysis showed that only 62% of the static representation paths exist (they follow a chronological order), showing that when comparing the static and the temporal representation there is a lack of a causal sequence of edges in the last, overestimating the number of transmission paths. These results are in line with the German pig network with a proportion of 74% (Lentz et al., 2016) but are opposite to the proportion of 11% in Santa Catarina, Brazil (Cardenas et al., 2021). We could associate this with the higher percentage of backyard production in Ecuador or the predominant commercialisation structure with markets and traders as important hubs. Further analysis is necessary to implement models considering the time variable in the simulations.

## 5. Conclusion

The characterisation of premises based on their contact risk has given relevant information to explore targeted surveillance and control strategies in a country with notable efforts to control CSF in the last few years but lacking formal studies on the swine industry to date.

The surveillance strategies applied by the National Veterinary Service could be more efficient in finding cases and reducing the spread of diseases if they focus their efforts on target selection of markets and traders that are evidenced to have similar disease spread potential. Additionally, we pay particular attention to other influential premises based on network centrality metrics. In a country dealing with a dichotomous swine producing sector, changes based on network modelling could enable the implementation of risk-based surveillance. Considering the temporal implications of the network and specifically targeting the nodes at higher risk, will reduce the time of the CSF eradication process and enable the prevention and control of other infectious diseases.

## 6. Data availability statement

The data on pig movements contain private information of each Ecuadorian pig farmer, therefore cannot be made available due to legal restrictions.

## 7. Funding

This work was supported by the São Paulo Research Foundation (FAPESP), grant 2017/22912-2.

## 8. Author contributions

Designed the study: AA MA CI FF. Performed database consolidation: CI AA. Processed the data: AA NC HL. Interpreted the results and reviewed for critical content: KD AB VG. All authors wrote the paper, discussed the results and critically reviewed the final manuscript.

## 9. Declaration of interests

The authors declare that they have no known competing financial interests or personal relationships that could have appeared to influence the work reported in this paper.

## 10. Ethics approval and consent to participate
Not applicable.

## 11. Acknowledgements

The authors would like to acknowledge the provision of data and support of the Phyto-zoosanitary Regulation and Control Agency of the Ministry of Agriculture of Ecuador (Agrocalidad), veterinarians and technical personnel at headquarters, and in their 23 provinces.

## 12. References

Acosta, A., Pisuna, L., Vasquez, S., Ferreira, F., 2019. Space-time cluster analysis to improve classical swine fever risk-based surveillance in Ecuador. Front. Vet. Sci. 6. https://doi.org/10.3389/conf.fvets.2019.05.00108

Acosta, A., Vargas, J., 2012. Proyecto de control y erradicación de la Peste porcina clásica por zonificación en el Ecuador. Agencia ecuatoriana de aseguramiento de la calidad del Agro-AGROCALIDAD, Ministerio de Agricultura, Ganaderia, Acuacultura y Pesca-MAGAP, Quito-Ecuador.

Alarcón, L.V., Cipriotti, P.A., Monterubbianessi, M., Perfumo, C., Mateu, E., Allepuz, A., 2020. Network analysis of pig movements in Argentina: Identification of key farms in the spread of infectious diseases and their biosecurity levels. Transboundary and Emerging Diseases 67, 1152–1163. https://doi.org/10.1111/tbed.13441

Andraud, M., Rose, N., 2020. Modelling infectious viral diseases in swine populations: a state of the art. Porc Health Manag 6, 22. https://doi.org/10.1186/s40813-020-00160-4

Baquero, O.S., 2019. ggsn: North Symbols and Scale Bars for Maps Created with "ggplot2" or "ggmap" [WWW Document]. URL https://CRAN.R-project.org/package=ggsn (accessed 1.1.20).

Baron, J.N., Aznar, M.N., Monterubbianesi, M., Martínez-López, B., 2020. Application of network analysis and cluster analysis for better prevention and control of swine diseases in Argentina. PLOS ONE 15, e0234489. https://doi.org/10.1371/journal.pone.0234489

Barongo, M.B., Bishop, R.P., Fèvre, E.M., Knobel, D.L., Ssematimba, A., 2016. A Mathematical Model that Simulates Control Options for African Swine Fever Virus (ASFV). PLOS ONE 11, e0158658. https://doi.org/10.1371/journal.pone.0158658

Bates, D., Maechler, M., 2021. Matrix: Sparse and Dense Matrix Classes and Methods [WWW Document]. URL https://CRAN.R-project.org/package=Matrix (accessed 7.5.21).

Beltran-Alcrudo, D., Falco, J.R., Raizman, E., Dietze, K., 2019. Transboundary spread of pig diseases: the role of international trade and travel. BMC Veterinary Research 15, 64–64. https://doi.org/10.1186/s12917-019-1800-5

Benitez, W., Sanchez, M.D., 2001. Los cerdos locales en los sistemas tradicionales de producción, 1st ed, Estudio FAO produccion y sanidad animal 148. Organización de las Naciones Unidas para la Agricultura y la Alimentación, Roma.

Bigras-Poulin, M., Thompson, R.A., Chriel, M., Mortensen, S., Greiner, M., 2006. Network analysis of Danish cattle industry trade patterns as an evaluation of risk potential for disease spread. Preventive Veterinary Medicine 76, 11–39. https://doi.org/10.1016/j.prevetmed.2006.04.004

Bivand, R., Keitt, T., Rowlingson, B., 2020. rgdal: Bindings for the "Geospatial" Data Abstraction Library [WWW Document]. URL https://CRAN.R-project.org/package=rgdal (accessed 9.14.20).

Bivand, R.S., Pebesma, E., Gomez-Rubio, V., 2013. Applied spatial data analysis with R, Second edition. Springer, NY.

Cardenas, N.C., Galvis, J.O.A., Farinati, A.A., Grisi-Filho, J.H.H., Diehl, G.N., Machado, G., 2019. Burkholderia mallei : The dynamics of networks and disease transmission. Transboundary and Emerging Diseases 66, 715–728. https://doi.org/10.1111/tbed.13071

Cardenas, N.C., VanderWaal, K., Veloso, F.P., Galvis, J.O.A., Amaku, M., Grisi-Filho, J.H.H., 2021. Spatio-temporal network analysis of pig trade to inform the design of risk-based disease surveillance. Preventive Veterinary Medicine 189, 105314. https://doi.org/10.1016/j.prevetmed.2021.105314

Craft, M.E., 2015. Infectious disease transmission and contact networks in wildlife and livestock. Phil. Trans. R. Soc. B 370, 20140107. https://doi.org/10.1098/rstb.2014.0107

Csardi, G., Nepusz, T., 2006. The igraph software package for complex network research [WWW Document]. InterJournal. URL https://igraph.org (accessed 9.1.20).

Dubé, C., Ribble, C., Kelton, D., McNab, B., 2009. A Review of Network Analysis Terminology and its Application to Foot-and-Mouth Disease Modelling and Policy Development. Transboundary and Emerging Diseases 56, 73–85. https://doi.org/10.1111/j.1865-1682.2008.01064.x

Edwards, S., Fukusho, A., Lefèvre, P.-C., Lipowski, A., Pejsak, Z., Roehe, P., Westergaard, J.J., Lefe Ávre, P.-C., Lipowski, A., Pejsak, Z., Roehe, P., Westergaard, J.J., 2000. Classical swine fever: the global situation. Veterinary Microbiology 73, 103–119. https://doi.org/10.1016/S0378-1135(00)00138-3

Encyclopedia Britannica, 2021. Ecuador - Agriculture, forestry, and fishing [WWW Document]. Encyclopedia Britannica. URL https://www.britannica.com/place/Ecuador (accessed 6.11.21).

Farine, D.R., Whitehead, H., 2015. Constructing, conducting and interpreting animal social network analysis. Journal of Animal Ecology 84, 1144–1163. https://doi.org/10.1111/1365-2656.12418

Frössling, J., Nusinovici, S., Nöremark, M., Widgren, S., Lindberg, A., 2014. A novel method to identify herds with an increased probability of disease introduction due to animal trade. Preventive Veterinary Medicine, Special Issue: SVEPM 2014 - supporting decision making on animal health through advanced and multidisciplinary methodologies, 2014 Society of Veterinary Epidemiology and Preventive Medicine conference 117, 367–374. https://doi.org/10.1016/j.prevetmed.2014.07.013

Ganges, L., Crooke, H.R., Bohórquez, J.A., Postel, A., Sakoda, Y., Becher, P., Ruggli, N., 2020. Classical swine fever virus: the past, present and future. Virus Research 289, 198151. https://doi.org/10.1016/j.virusres.2020.198151

Guinat, C., Relun, A., Wall, B., Morris, A., Dixon, L., Pfeiffer, D.U., 2016. Exploring pig trade patterns to inform the design of risk-based disease surveillance and control strategies. Nature Publishing Group. https://doi.org/10.1038/srep28429

Hagberg, A.A., Schult, D.A., Swart, P.J., 2008. Exploring Network Structure, Dynamics, and Function using NetworkX [WWW Document]. Proceedings of the 7th Python in Science Conference. URL https://conference.scipy.org/proceedings/SciPy2008/paper_2/ (accessed 5.27.21).

Harris, C.R., Millman, K.J., van der Walt, S.J., Gommers, R., Virtanen, P., Cournapeau, D., Wieser, E., Taylor, J., Berg, S., Smith, N.J., Kern, R., Picus, M., Hoyer, S., van Kerkwijk, M.H., Brett, M., Haldane, A., del Río, J.F., Wiebe, M., Peterson, P., Gérard-Marchant, P., Sheppard, K., Reddy, T., Weckesser, W., Abbasi, H., Gohlke, C., Oliphant, T.E., 2020. Array programming with NumPy. Nature 585, 357–362. https://doi.org/10.1038/s41586-020-2649-2

Hijmans, R.J., 2020. raster: Geographic Data Analysis and Modeling [WWW Document]. URL https://CRAN.R-project.org/package=raster (accessed 8.1.20).

Holme, P., Saramäki, J., 2012. Temporal networks. Physics Reports, Temporal Networks 519, 97–125. https://doi.org/10.1016/j.physrep.2012.03.001

James, R., Croft, D.P., Krause, J., 2009. Potential banana skins in animal social network analysis. Behav Ecol Sociobiol 63, 989–997. https://doi.org/10.1007/s00265-009-0742-5

Kahle, D., Wickham, H., 2013. ggmap: Spatial Visualization with ggplot2 [WWW Document]. The R Journal. URL https://journal.r-project.org/archive/2013-1/kahle-wickham.pdf

Keeling, M.J., Danon, L., 2009. Mathematical modelling of infectious diseases. British Medical Bulletin 92, 33–42. https://doi.org/10.1093/bmb/ldp038

Keeling, M.J., Rohani, Pejman., 2011. Modeling Infectious Diseases in Humans and Animals. Princeton University Press.

Kempe, D., Kleinberg, J., Kumar, A., 2002. Connectivity and Inference Problems for Temporal Networks. Journal of Computer and System Sciences 64, 820–842. https://doi.org/10.1006/jcss.2002.1829

Knific, T., Ocepek, M., Kirbiš, A., Lentz, H.H.K., 2020. Implications of Cattle Trade for the Spread and Control of Infectious Diseases in Slovenia. Frontiers in Veterinary Science 6, 454. https://doi.org/10.3389/fvets.2019.00454

Koeppel, L., Siems, T., Fischer, M., Lentz, H.H.K., 2018. Automatic classification of farms and traders in the pig production chain. Preventive Veterinary Medicine 150, 86–92. https://doi.org/10.1016/j.prevetmed.2017.12.003

Lentz, H.H.K., Selhorst, T., Sokolov, I.M., 2013. Unfolding Accessibility Provides a Macroscopic Approach to Temporal Networks. Phys. Rev. Lett. 110, 118701. https://doi.org/10.1103/PhysRevLett.110.118701

Lentz, Koher, A., Hövel, P., Gethmann, J., Sauter-Louis, C., Selhorst, T., Conraths, F.J., 2016. Disease spread through animal movements: A static and temporal network analysis of pig trade in Germany. PLoS ONE 11, e0155196–e0155196. https://doi.org/10.1371/journal.pone.0155196

Lloyd-Smith, J.O., Schreiber, S.J., Kopp, P.E., Getz, W.M., 2005. Superspreading and the effect of individual variation on disease emergence. Nature 438, 355–359. https://doi.org/10.1038/nature04153

Makau, D.N., Paploski, I.A.D., VanderWaal, K., 2021. Temporal stability of swine movement networks in the U.S. Preventive Veterinary Medicine 191, 105369. https://doi.org/10.1016/j.prevetmed.2021.105369

Mangen, M.-J.J., Nielen, M., Burrell, A.M., 2002. Simulated effect of pig-population density on epidemic size and choice of control strategy for classical swine fever epidemics in The Netherlands. Preventive Veterinary Medicine 56, 141–163. https://doi.org/10.1016/S0167-5877(02)00155-1

Masuda, N., Holme, P., 2013. Predicting and controlling infectious disease epidemics using temporal networks. F1000Prime Rep 5. https://doi.org/10.12703/P5-6

May, R.M., Lloyd, A.L., 2001. Infection dynamics on scale-free networks. Phys. Rev. E 64, 066112. https://doi.org/10.1103/PhysRevE.64.066112

Motta, P., Porphyre, T., Handel, I., Hamman, S.M., Ngu Ngwa, V., Tanya, V., Morgan, K., Christley, R., Bronsvoort, B.M. deC, 2017. Implications of the cattle trade network in Cameroon for regional disease prevention and control. Sci Rep 7, 43932.
https://doi.org/10.1038/srep43932

National Research Council, 1991. Microlivestock: Little-Known Small Animals with a Promising Economic Future. The National Academies Press, Washington, DC. https://doi.org/10.17226/1831

Newman, M., 2008. The physics of networks. Physics Today 61, 33–38. https://doi.org/10.1063/1.3027989

Newman, M.E.J. (Mark E.J.), 2010. Networks : an introduction. Oxford University Press.

OIE, 2021a. Terrestrial Manual Online Access [WWW Document]. OIE - World Organisation for Animal Health. URL https://www.oie.int/en/what-we-do/standards/codes-and-manuals/terrestrial-manual-online-access/ (accessed 2.14.21).

OIE, 2021b. Resolution No. 20 [WWW Document]. URL https://www.oie.int/app/uploads/2021/05/a-r20-2021-csf.pdf (accessed 7.15.21).

Ossada, R., Grisi-Filho, J.H.H., Ferreira, F., Amaku, M., 2013. Modeling the Dynamics of Infectious Diseases in Different Scale-Free Networks with the Same Degree Distribution. Adv. Studies Theor. Phys 7, 759–771. https://doi.org/10.12988/astp.2013.3674

Pineda, P., Deluque, A., Peña, M., Diaz, O.L., Allepuz, A., Casal, J., 2020. Descriptive epidemiology of classical swine fever outbreaks in the period 2013-2018 in Colombia. PLOS ONE 15, e0234490. https://doi.org/10.1371/journal.pone.0234490

R Core Team, 2020. R: A Language and Environment for Statistical Computing [WWW Document]. URL https://www.r-project.org/ (accessed 10.1.21).

Robinson, S.E., Christley, R.M., 2007. Exploring the role of auction markets in cattle movements within Great Britain. Preventive Veterinary Medicine 81, 21–37. https://doi.org/10.1016/j.prevetmed.2007.04.011

Silk, M.J., Croft, D.P., Delahay, R.J., Hodgson, D.J., Boots, M., Weber, N., McDonald, R.A., 2017. Using Social Network Measures in Wildlife Disease Ecology, Epidemiology, and Management. BioScience 67, 245–257. https://doi.org/10.1093/biosci/biw175

Stegeman, J., Elbers, A.R.W., Bouma, A., Jong, M.C.M., 2002. Rate of inter-herd transmission of classical swine fever virus by different types of contact during the 1997–8 epidemic in The Netherlands. Epidemiol. Infect 128, 285–291. https://doi.org/10.1017/s0950268801006483

Terán, M.V., Ferrat, N.C., Lubroth, J., 2004. Situation of Classical Swine Fever and the Epidemiologic and Ecologic Aspects Affecting Its Distribution in the American Continent. Annals of the New York Academy of Sciences 1026, 54–64. https://doi.org/10.1196/annals.1307.007

Valverde Lucio, A., Gonzalez-Martínez, A., Alcívar Cobeña, J.L., Rodero Serrano, E., 2021. Characterization and Typology of Backyard Small Pig Farms in Jipijapa, Ecuador. Animals 11, 1728. https://doi.org/10.3390/ani11061728

Virtanen, P., Gommers, R., Oliphant, T.E., Haberland, M., Reddy, T., Cournapeau, D., Burovski, E., Peterson, P., Weckesser, W., Bright, J., van der Walt, S.J., Brett, M., Wilson, J., Millman, K.J., Mayorov, N., Nelson, A.R.J., Jones, E., Kern, R., Larson, E., Carey, C.J., Polat, İ., Feng, Y., Moore, E.W., VanderPlas, J., Laxalde, D., Perktold, J., Cimrman, R., Henriksen, I., Quintero, E.A., Harris, C.R., Archibald, A.M., Ribeiro, A.H., Pedregosa, F., van Mulbregt, P., 2020. SciPy 1.0: fundamental algorithms for scientific computing in Python. Nature Methods 17, 261–272. https://doi.org/10.1038/s41592-019-0686-2

Warnes, G.R., Bolker, B., Gorjanc, G., Grothendieck, G., Korosec, A., Lumley, T., MacQueen, D., Magnusson, A., Rogers, J., others, 2017. gdata: Various R Programming Tools for Data Manipulation [WWW Document]. URL https://CRAN.R-project.org/package=gdata (accessed 2.1.18).

Wasserman, S., Urbana-Champaign), S. (University of I.W., Faust, K., 1994. Social Network Analysis: Methods and Applications. Cambridge University Press.

Wickham, H., Averick, M., Bryan, J., Chang, W., McGowan, L.D., François, R., Grolemund, G., Hayes, A., Henry, L., Hester, J., Kuhn, M., Pedersen, T.L., Miller, E., Bache, S.M., Müller, K., Ooms, J., Robinson, D., Seidel, D.P., Spinu, V., Takahashi, K., Vaughan, D., Wilke, C., Woo, K., Yutani, H., 2019. Welcome to the Tidyverse. Journal of Open Source Software 4, 1686. https://doi.org/10.21105/joss.01686

**Modelling control strategies against Classical Swine Fever: influence of traders and markets using static and temporal networks in Ecuador**

Alfredo Acosta [1], Nicolas Cespedes Cardenas[3], Cristian Imbacuan[2], Hartmut H.K. Lentz[4], Klaas Dietze[4], Marcos Amaku[1], Alexandra Burbano[2] Vitor S.P. Gonçalves[5], Fernando Ferreira[1]

1 Preventive Veterinary Medicine Department, School of Veterinary Medicine and Animal Science, University of São Paulo, SP, Brazil
2 Sanitary and phytosanitary control and regulation Agency-Agrocalidad, Ministry of Agriculture, Quito, Ecuador
3 Department of Population Health and Pathobiology, College of Veterinary Medicine, North Carolina State University, Raleigh, USA
4 Friedrich-Loeffler-Institut, Greifswald-Riems, Germany
5 Faculty of Agronomy and Veterinary Medicine, University of Brasilia, Brasilia, Brazil

Corresponding author: Alfredo Acosta, Preventive Veterinary Medicine Department, School of Veterinary Medicine and Animal Science. University of São Paulo.

alfredojavier55@gmail.com, alfredoacosta@usp.br

**Supplementary materials (SM)**

Table SM1. Market centrality metrics ordered by their degree value.

| Market name | pagerank | Degree | Weighted Degree (pigs) | Quintile | Mean Degree by quintile | Mean Weighted degree (pigs) by quintile | Province |
|---|---|---|---|---|---|---|---|
| Asogansd | 0.0274 | 29,407 | 241,572 | 5 | 13,628 | 59,684 | Santo Domingo Tsáchilas |
| Saquisili | 0.0309 | 24,727 | 64,285 | 5 | 13,628 | 59,684 | Cotopaxi |
| Salcedo | 0.0404 | 18,856 | 64,128 | 5 | 13,628 | 59,684 | Cotopaxi |
| Patamarca | 0.0063 | 5,518 | 59,269 | 4 | 4,610 | 20,221 | Azuay |
| Asoganec | 0.0058 | 6,796 | 42,982 | 5 | 13,628 | 59,684 | Manabí |
| La cruz | 0.0127 | 10,346 | 37,495 | 5 | 13,628 | 59,684 | Imbabura |
| Sangolqui | 0.0088 | 10,013 | 32,203 | 5 | 13,628 | 59,684 | Pichincha |
| Machachi | 0.0094 | 8,185 | 30,964 | 5 | 13,628 | 59,684 | Pichincha |
| Montufar | 0.0095 | 9,142 | 28,596 | 5 | 13,628 | 59,684 | Carchi |
| Plateado | 0.0114 | 5,990 | 27,711 | 5 | 13,628 | 59,684 | Loja |
| Guanujo | 0.0299 | 12,815 | 26,909 | 5 | 13,628 | 59,684 | Bolívar |
| Ambato | 0.0132 | 5,355 | 26,557 | 4 | 4,610 | 20,221 | Tungurahua |

| | | | | | | |
|---|---|---|---|---|---|---|
| Zumbalica | 0.0065 | 4,611 | 19,397 | 4 | 4,610 | 20,221 Cotopaxi |
| Echeandia | 0.0052 | 4,454 | 17,924 | 4 | 4,610 | 20,221 Bolívar |
| Riobamba | 0.0134 | 3,650 | 17,389 | 3 | 1,948 | 7,774 Chimborazo |
| Quero | 0.0109 | 4,560 | 15,987 | 4 | 4,610 | 20,221 Tungurahua |
| Otavalo | 0.0035 | 3,767 | 15,551 | 4 | 4,610 | 20,221 Imbabura |
| Cayambe | 0.0038 | 4,159 | 13549 | 4 | 4,610 | 20,221 Pichincha |
| Pucayacu | 0.0033 | 2,505 | 12,721 | 3 | 1,948 | 7,774 Cotopaxi |
| Pelileo | 0.0071 | 3,040 | 12,216 | 3 | 1,948 | 7,774 Tungurahua |
| Pillaro | 0.0106 | 5,334 | 12,123 | 4 | 4,610 | 20,221 Tungurahua |
| J andrade | 0.0042 | 4,399 | 11,666 | 4 | 4,610 | 20,221 Carchi |
| Tulcan | 0.0059 | 3,941 | 10,184 | 4 | 4,610 | 20,221 Carchi |
| Calderon | 0.0017 | 2,022 | 7,355 | 3 | 1,948 | 7,774 Pichincha |
| Sm bolivar | 0.0070 | 2,522 | 6,785 | 3 | 1,948 | 7,774 Bolívar |
| Pallatanga | 0.0024 | 1,311 | 5,848 | 3 | 1,948 | 7,774 Chimborazo |
| P maldonado | 0.0008 | 1,072 | 5,569 | 3 | 1,948 | 7,774 Pichincha |
| Cevallos | 0.0040 | 1,703 | 4,934 | 3 | 1,948 | 7,774 Tungurahua |
| Biblian | 0.0021 | 713 | 3,935 | 2 | 456 | 1,795 Cañar |
| Indanza | 0.0013 | 933 | 3,133 | 3 | 1,948 | 7,774 Morona Santiago |
| Guamote | 0.0030 | 601 | 2,961 | 2 | 456 | 1,795 Chimborazo |
| Yantzaza | 0.0010 | 447 | 2,440 | 2 | 456 | 1,795 Zamora Chinchipe |
| Gualaceo | 0.0009 | 364 | 2,108 | 2 | 456 | 1,795 Azuay |
| Chunchi | 0.0022 | 720 | 1,792 | 3 | 1,948 | 7,774 Chimborazo |
| Pujili | 0.0014 | 703 | 1,762 | 2 | 456 | 1,795 Cotopaxi |
| R del cisne | 0.0005 | 233 | 1,662 | 2 | 456 | 1,795 Orellana |
| Sucumbios | 0.0002 | 114 | 1,098 | 1 | 84 | ,395 Sucumbíos |
| Alausi | 0.0038 | 509 | 1,045 | 2 | 456 | 1,795 Chimborazo |
| Chillanes | 0.0021 | 477 | 1,006 | 2 | 456 | 1,795 Bolívar |
| Santa isabel | 0.0006 | 165 | 762 | 1 | 84 | 395 Azuay |
| Caluma | 0.0006 | 224 | 645 | 2 | 456 | 1,795 Bolívar |
| Huambalo | 0.0003 | 167 | 631 | 1 | 84 | 395 Tungurahua |
| Colta | 0.0003 | 52 | 492 | 1 | 84 | 395 Chimborazo |
| Simiatug | 0.0011 | 291 | 383 | 2 | 456 | 1,795 Bolívar |
| Paute | 0.0004 | 92 | 383 | 1 | 84 | 395 Azuay |
| Sigchos | 0.0005 | 121 | 323 | 1 | 84 | 395 Cotopaxi |
| Cumbe | 0.0002 | 42 | 217 | 1 | 84 | 395 Azuay |

| San lucas | 0.0002 | 94 | 187 | 1 | 84 | 395 Loja |
| Moraspungo | 0.0001 | 23 | 110 | 1 | 84 | 395 Cotopaxi |
| Saraguro | 0.0001 | 37 | 96 | 1 | 84 | 395 Loja |
| Canar | 0.0001 | 13 | 52 | 1 | 84 | 395 Cañar |

Fig. SM2. Market box plots of pagerank distribution, classified by quintile of degree in Ecuador

(Markets' original names in Spanish) Average annual values in the study period.

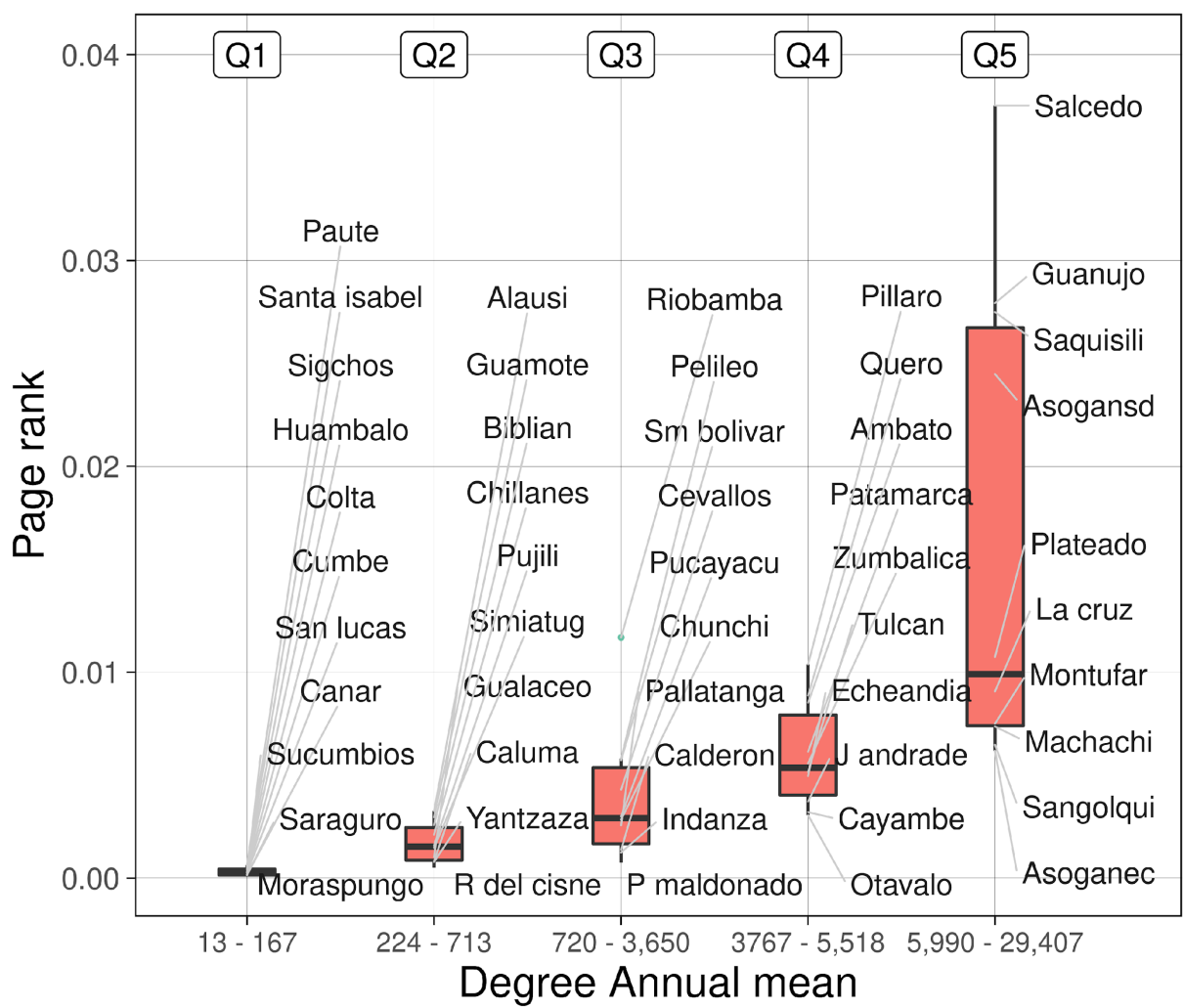